\documentclass[12pt]{article}

\usepackage{amsmath,amssymb,a4wide}
\usepackage{graphicx}
\numberwithin{equation}{section}

\title{\bf Scaling prediction for self-avoiding \\
polygons revisited}

\author{\sc C.~Richard$^{1)}$, I.~Jensen$^{2)}$ and A.~J.~Guttmann$^{2)}$\\
\\
${}^{1)}$Fakult\"at f\"ur Mathematik, Universit\"at Bielefeld\\
Postfach 10 01 31, 33501 Bielefeld, Germany\\
${}^{2)}$ARC Centre of Excellence \\
for the Mathematics and Statistics of Complex Systems\\
Department of Mathematics and Statistics \\
The University of Melbourne, Victoria 3010, Australia
}

\begin{document}

\maketitle

\begin{abstract}
We analyse new exact enumeration data for self-avoiding polygons, counted by
perimeter and area on the square, triangular and hexagonal lattices. In 
extending earlier analyses, we focus on the perimeter moments in the vicinity 
of the bicritical point. We also consider the shape of the critical curve near the
bicritical point, which describes the crossover to the branched polymer phase.
Our recently conjectured expression for the scaling function of rooted self-avoiding 
polygons is further supported. For (unrooted) self-avoiding polygons, the analysis 
reveals the presence of an additional additive term with a new universal amplitude. 
We conjecture the exact value of this amplitude. 
\end{abstract}

\section{Introduction}

The model of (planar) self-avoiding polygons \cite{MS93, JvR00} is an important 
unsolved model of statistical physics. Some progress has been made in recent years.
Results from exactly solvable polygon models led to a prediction of the scaling 
function of self-avoiding polygons, counted by perimeter and area \cite{RGJ01, C01}. 
On the other hand, the theory of stochastic processes provided new insight into 
the problem by relating it to the so-called Schramm-Loewner evolution \cite{LSW02}. 
In this article, we will further test the predictions implied by the scaling 
function conjecture \cite{RGJ01,C01} and extend it.

Let $p_{m,n}$ denote the number of self-avoiding polygons (SAP) of
perimeter $m$ and area $n$ on a given lattice. In this article, we will
consider self-avoiding polygons on the square, hexagonal and triangular 
lattices. Denote the perimeter and area generating function by 
$G(x,q)=\sum_{m,n} p_{m,n} x^m q^n$.  The function $G(x,1)$ is called the 
perimeter generating function of the model, its radius of convergence is denoted 
by $x_c$. For SAP, we have for the singular part of the perimeter generating function 
$G^{(sing)}(x,1) \sim A(1 - x/x_c)^{2 - \alpha}$ as $x\to x_c$, with $\alpha = 1/2$ being 
universally accepted, though not rigorously proved. (All limits appearing in this paper
will be taken from below.) Until section 5, we will concentrate on {\em rooted} 
self-avoiding polygons, whose perimeter and area generating function is 
$G^{(r)}(x,q)=\sum_{m,n} p^{(r)}_{m,n} x^m q^n$, where $p^{(r)}_{m,n}=mp_{m,n}$.
We thus have $G^{(r)}(x,q)=x\frac{\rm d}{{\rm d}x}G(x,q)$
and $G^{(r),(sing)}(x,1) \sim B(1 - x/x_c)^{1 - \alpha}$ as $x\to x_c$.

The phase diagram of SAP, enumerated by perimeter and area, appears to have first
been discussed in \cite{FGW}. There is a phase boundary in the region $q < 1,$ $x > x_c,$
terminating in a bicritical point at $(x_c,1)$. If $q=1$, typical polygons are extended, 
whereas for $q<1$, typical polygons try to minimise their area. This particular type of
phase transition is also called a {\em collapse transition}, and the phase $q<1$ is called the 
branched polymer phase. Indeed, polygons of minimal area may be viewed as branched polymers. 
The phase boundary is characterised by a logarithmic singularity, when approached from below. 
This particular feature was first found by studying the area generating function of SAP, 
$G(1,q)$, which was found (numerically) to have a singularity of the form 
$G(1,q) \sim A\log(1 - q)$ as $q\to1$ \cite{EG90}.
The point $(x,q)=(x_c,1)$ is a bicritical point where a generic scaling form of the 
perimeter and area generating function is expected to hold \cite{FGW}. The singular behaviour 
about a bicritical point is generally expected to be of the form
\begin{equation}\label{form:scal}
G(x,q) \sim G^{(reg)}(x,q) + (1-q)^\theta H\left(\frac{x_c-x}{(1-q)^\phi}\right) 
\qquad (x,q) \to (x_c,1),
\end{equation}
where $G^{(reg)}(x,q)$ denotes the regular part of $G(x,q)$ at $(x,q)=(x_c,1)$, 
and $H(s)$ is called the scaling function with critical exponents $\theta$ and $\phi$,  
and $s = \frac{x_c-x}{(1-q)^\phi}$. We stress that there are counter-examples known where 
such a scaling form is not valid, for example, in the simple model of rectangles \cite{JvR04}.
For staircase polygons \cite{P95}, the behaviour (\ref{form:scal}) has been proved. The exponents 
are $\theta=1/3$, $\phi=2/3$, and the scaling function of staircase polygons is the 
logarithmic derivative of an Airy function. The phase diagram of staircase polygons is 
similar to that of SAP. There is also a phase boundary in the region $q < 1,$ $x > x_c$, 
terminating in a bicritical point at $(x_c,1)$. The phase boundary in that case is characterised 
by a simple pole, when approached from below, while at the bicritical point we have a branch-point 
singularity $G^{(sing)}(x_c,q) \sim B(1 - q)^{1/3}$ as $q\to1$, as follows from \cite{P95}. 
Interestingly, {\em rooted} SAPs display the same singularity structure as the staircase polygon
model. This led to the question whether the scaling functions might be the same \cite{PO95}.

Any conjectured form for the singular behaviour of $G^{(r)}(x,q)$ can most appropriately be 
tested by comparing predicted moments to those calculated numerically. In \cite{J03,RGJ01} we 
made such a comparison with the {\em area} moments, which led us to conjecture the exact form
of the scaling function, thereby answering the previous question in the affirmative. We review 
this calculation below. The validity of the scaling function conjecture leads, in addition, 
to predictions of the leading singular behaviour of the {\em perimeter} moments, as explained 
in section 3. Checking this behaviour thus yields a further test of the scaling assumption.
This was done in \cite{RGJ01} only for the moment of order zero, i.e., the 
(bicritical) area generating function $G^{(r)}(x_c,q)$ as $q\to1$. Here we present, 
for the first time, a detailed numerical analysis for the higher moments, which turns out to be 
numerically much more difficult than the analysis of the area moments. 

After establishing agreement with the predictions of the values of the first 10 perimeter 
moments in section 4, we consider the scaling functions for (unrooted) SAP, obtained
by integration of the rooted SAP scaling function. The ``constant'' of integration,
which must be a function solely of $q$, in order that its derivative with respect to
$x$ vanishes, turns out to dominate the behaviour of the scaling function
as $q \rightarrow 1.$ We argue for a particular form of this term, and then numerical
testing reveals an unexpected amplitude universality across the three lattices we
study. Based on our experience with other exact amplitudes, we conjecture the exact
value of this universal amplitude. More precisely, we show below that at the bicritical point 
the behaviour is $G^{(sing)}(x_c,q) \sim A(1 - q)\log(1 - q)$ as $q\to1$, where we 
conjecture the exact value of the amplitude $A$. Our findings imply that the scaling form
(\ref{form:scal}) cannot hold for (unrooted) SAP. We suggest a modified form below, see 
(\ref{form:newscal}), (\ref{form:oldsf}) and (\ref{form:add}).

In the last section, we analyse the shape of the critical curve near the bicritical point, 
which describes the crossover to the branched polymer phase. We find the prediction from 
the scaling function conjecture satisfied, within numerical accuracy. We also analyse the
behaviour of the critical curve as $q\to0$. Numerical techniques are explained in an appendix.

\section{Area moments for rooted SAP}

The factorial area moment generating functions $g^{(r)}_k(x)$
of rooted self-avoiding polygons are defined by
\begin{equation}\label{form:armom}
g^{(r)}_k(x) = \frac{(-1)^k}{k!} \left. \frac{{\rm d}^k}{{\rm d}q^k} G^{(r)}(x,q)
\right|_{q=1}
= \frac{(-1)^k}{k!} \sum_{m,n} (n)_k \, p^{(r)}_{m,n} x^m,
\end{equation}
where $(a)_k=a(a-1)\cdots (a-k+1)$.  Previous numerical analyses \cite{J03,RGJ01}
based on exact enumeration data provide strong evidence for the asymptotic form
\begin{equation}\label{form:armomas}
g^{(r),(sing)}_k(x) \sim \frac{f_k}{(x_c-x)^{\gamma_k}} \qquad (x\to x_c),
\end{equation}
with exponents $\gamma_k=(3k-1)/2$.  We incorporate the coefficients and exponents 
into the  function
$F^{(r)}(s)$  defined by
\begin{equation}\label{form:scall}
F^{(r)}(s) = \sum_{k=0}^\infty \frac{f_k}{s^{\gamma_k}}.
\end{equation}
At this stage, the function $F^{(r)}(s)$ should be viewed as some generating function
for the numbers $f_k$, with $s$ being an undetermined variable. We will argue below that, 
given the validity of the scaling assumption (\ref{form:scal}), the function $F^{(r)}(s)$ 
is the scaling function $H^{(r)}(s)$ of the model.%
\footnote{
The superscript $\mbox{}^{(r)}$ indicates the relation to rooted SAP.}
(In fact, $F^{(r)}(s)$ is then the generating function of the perimeter moment 
amplitudes $e_k$, which are defined in (\ref{form:permomas})).

In \cite{RGJ01}, we tested the conjecture that 
the function $F^{(r)}(s)$ of rooted self-avoiding polygons is given by
\begin{equation}\label{form:sfun}
F^{(r)}(s)=-4 f_1 \frac{d}{ds}\log\mbox{Ai}\left( \left(
\frac{f_0}{4f_1}\right)^{2/3}s\right)
\end{equation}
by comparing numerical estimates of $f_k$ with estimates that follow from (\ref{form:sfun}).
The constants $f_0$ and $f_1$ have been numerically determined previously to great 
accuracy \cite{RJG03}. We have $f_0 = -\frac{2 E_0\sqrt{\pi}}{\sigma \sqrt{x_c}}$ and
$f_1 = -\frac{E_1 x_c}{\sigma}$, where $x_c$ is the radius of convergence of the 
perimeter generating function, given by $x_c=0.379052277757(5)$ on the square lattice, 
$x_c=0.2409175745(3)$ on the triangular lattice, and $x_c=1/\sqrt{2+\sqrt2}$ on the 
hexagonal lattice \cite{N82}. The constant $\sigma$ is defined such that $p_{m,n}$ is 
nonzero if $m$ is divisible by $\sigma$. Thus $\sigma=2$ for the square and hexagonal 
lattices and $\sigma=1$ for the triangular lattice. Estimates of the amplitudes $E_0$ 
and $E_1$, taken from \cite{RJG03}, are given in Table \ref{tab:amp}. The value 
$E_1=1/(4\pi)=0.07957747\ldots$ has been derived in \cite{C94}, using field 
theoretic arguments.
\begin{table}
\begin{center}
\begin{tabular}{cccc}
Amplitude & Square & Hexagonal & Triangular\\ \hline
$E_0$ & $0.56230130(2)$ & $1.27192995(10)$ & $0.2639393(2)$\\
$E_1$ & $0.0795773(2)$ & $0.0795779(5)$ & $0.0795765(10)$ \\
$f_0$ & $-1.61880474(6)$ & $-3.0645083(3)$ & $-1.906228(1)$ \\
$f_1$ & $-0.01508198(4)$ & $-0.0215332(3)$ & $-0.0191714(2)$
\end{tabular}
\end{center}
\caption{\label{tab:amp} Values of area moment amplitudes for SAPs taken from \cite{RJG03}}
\end{table}
The conjecture (\ref{form:sfun}) follows from the assumption that rooted self-avoiding 
polygons behave asymptotically like models whose perimeter and area generating function 
are described by a $q$-algebraic functional equation of arbitrary degree with a square-root singularity as 
the dominant singularity of the perimeter generating function \cite{D99,RGJ01,R02}. This class
includes a number of exactly solved polygons models such as staircase polygons, 
column-convex polygons, and bar-graph polygons \cite{PB95,JvR00}. For models within this class, 
the coefficients $f_k$ and exponents $\gamma_k$ from (\ref{form:armomas}) have been explicitly
calculated, leading to the expression (\ref{form:sfun}). It is interesting to note 
that the distribution of area in the limit of large perimeter, which can be extracted from 
(\ref{form:sfun}), see also \cite[Appendix]{R04}, is given by the Airy distribution, which 
appears in a number of related contexts \cite{FL01, R04}. 

Assuming the validity of the scaling form (\ref{form:scal}), the singular behaviour of the area 
moment generating functions (\ref{form:armomas}) determines the critical exponents and the 
scaling function $H^{(r)}(s)$. Taking the limit $q\to1$ in (\ref{form:scal}), we infer from 
(\ref{form:armomas}), leaving $\gamma_k$ unspecified for the moment, that
\begin{equation}
\gamma_k = \frac{k-\theta}{\phi}.
\end{equation}
We also find that the asymptotic expansion of the scaling function $H^{(r)}(s)$ is 
given by $H^{(r)}(s)=F^{(r)}(s)$, where $F^{(r)}(s)$ is defined in (\ref{form:scall}).
Thus, for rooted SAP, the assumption of the scaling form (\ref{form:scal}), together with the 
result (\ref{form:sfun}) for the area moments, leads to exponents $\theta=1/3$ and $\phi=2/3$ 
and to the scaling function $H^{(r)}(s)=F^{(r)}(s)$, where $F^{(r)}(s)$ is given in 
(\ref{form:sfun}).

\section{Perimeter moments}

The factorial perimeter moment generating functions $h^{(r)}_k(q)$ of rooted 
self-avoiding polygons at the bicritical point are defined by
\begin{equation}\label{form:permom}
h^{(r)}_k(q) = \frac{(-1)^k}{k!} \left. \frac{{\rm d}^k}{{\rm d}x^k} G^{(r)}(x,q)
\right|_{x=x_c}
= \frac{(-1)^k}{k!} \sum_{m,n} (m)_k \, p^{(r)}_{m,n} x_c^m q^n,
\end{equation}
The scaling assumption (\ref{form:scal}) leads to a prediction for the behaviour 
of the perimeter and area generating function in the limit $x\to x_c$. The leading
singular behaviour of the perimeter moments is given by
\begin{equation}\label{form:permomas}
h^{(r),(sing)}_k(q) \sim \frac{e_k}{(1-q)^{\beta_k}} \qquad (q\to 1),
\end{equation}
with exponents $\beta_k = k\phi-\theta$, where the singular amplitudes $e_k$ 
appear in the expansion of the scaling function $H^{(r)}(s)=F^{(r)}(s)$ about the origin,
\begin{equation}\label{form:perexpan}
F^{(r)}(s) = \sum_{k=0}^\infty e_k s^k.
\end{equation}
We can readily derive formulae for the expansion coefficients $e_k$ in (\ref{form:perexpan}). 
Note that the scaling function (\ref{form:sfun})
satisfies the Riccati equation
\begin{equation}\label{form:Ric}
F^{(r)}(s)^2-4 f_1 {F^{(r)}}'(s) -f_0^2=0.
\end{equation}
Inserting the form (\ref{form:perexpan}) into (\ref{form:Ric}), we obtain for 
the numbers $e_n$ the expression
\begin{equation}\label{form:par}
e_n=b_n f_1 \left(\frac{f_0}{f_1}\right)^{(2n+2)/3},
\end{equation}
where the constants $b_n$ are defined by the quadratic recursion
\begin{equation}
4 n b_n + \delta_{n-1,1}=\sum_{k=0}^{n-1} b_k b_{n-1-k} \qquad (n>0).
\end{equation}
The constants $b_n$ are polynomials of degree $n+1$ in $b_0$. The value
of $b_0$, remaining undetermined by (\ref{form:Ric}), can be extracted
from the limit $s\to0$ in (\ref{form:sfun}) as
\begin{equation}
b_0= \frac{3^\frac{5}{6}\Gamma(\frac{2}{3})^22^\frac{2}{3}}{2\pi }.
\end{equation}
It is interesting to note that the perimeter moments are related to
the area moments of negative order \cite[Eqn.~(37)]{FL01}.
It follows from (\ref{form:par}) that the amplitude combinations 
$e_k e_1^{-k} e_0^{k-1}=b_k b_1^{-k} b_0^{k-1}$ are independent of the 
amplitudes $f_0$ and $f_1$. The first few combinations are
\begin{equation}
\begin{split}
e_2 e_1^{-2} e_0 &=1-\frac{4}{27}\frac{\sqrt{3}
\pi^3}{\Gamma(\frac{2}{3})^6}, \qquad
e_3 e_1^{-3} e_0^{2} =1-\frac{8}{81}\frac{\sqrt{3}
\pi^3}{\Gamma(\frac{2}{3})^6},\\
e_4 e_1^{-4} e_0^{3} &=1-\frac{10}{81}\frac{\sqrt{3}
\pi^3}{\Gamma(\frac{2}{3})^6},\qquad
e_5 e_1^{-5} e_0^{4} =1-\frac{4}{27}\frac{\sqrt{3}
\pi^3}{\Gamma(\frac{2}{3})^6}+\frac{16}{1215}
\frac{\pi^6}{\Gamma(\frac{2}{3})^{12}}.
\end{split}
\end{equation}

In the next section we compare these predictions with the numerical values
obtained from our new enumeration data.

\section{Perimeter moment analysis}

We have generated data for self-avoiding polygons, counted by perimeter and area
on the square, hexagonal and triangular lattices, using the finite lattice method.
In particular, we determined the numbers $p_{m,n}$ for $n\le50$ (square lattice), 
$n\le40$ (hexagonal lattice) and $n\le60$ (triangular lattice), for all relevant
perimeter lengths. 
The algorithms used in our SAP enumerations are based on the finite-lattice method 
devised by Enting \cite{IGE80} in his pioneering work on the enumeration of
polygons on the square lattice. Details of the algorithms used to enumerate SAPs on
the hexagonal and triangular lattices can be found in \cite{EG89} and \cite{EG92},
respectively. A major enhancement, resulting in exponentially more efficient 
algorithms, is described in some detail in \cite{JG99} while recent work on
parallel versions can be found in \cite{J03}. All of the algorithms described in
these papers are for enumerations by perimeter, but the generalisation to include area 
is straightforward. The calculations were performed on the server cluster
of the Australian Partnership for Advanced Computing (APAC). The calculations
for the square lattice required up to 14Gb of memory, and were performed on up to 16
processors using a total of just under 2000 CPU hours. Comparable computational
resources were required for the hexagonal and triangular lattices.

We first checked the prediction for the exponents $\beta_k=2k/3-1/3$ defined below 
(\ref{form:permomas}), using first order differential approximants \cite{G89}. Then, we 
estimated the amplitudes $e_0$ and $e_1$ by a direct fit of the data to the expected 
asymptotic form, as explained in the appendix. Using the notation of the appendix, we 
fitted with exponents of the form $\alpha_i=(i+1)/3$. In the data analysis, we had 
$1\le i\le M_0$, where $2\le M_0\le4$. 

The particular choice of exponents $\alpha_i$ arises from the numerically well 
established behaviour of the area moment generating function
\begin{equation}
g^{(r)}_k(x) \sim \sum_{l=0}^L \frac{f_{k,l}}{(x_c-x)^{\gamma_{k,l}}} \qquad 
(x\to x_c),
\end{equation}
with exponents
\begin{equation}\label{form:generexp}
\gamma_{k,l}=(3k-l-1)/2,
\end{equation}
where $f_{k,1}=0$ and $f_{0,2l+1}=0$, see \cite{RGJ01}. If, in generalising 
(\ref{form:scal}), a scaling behaviour of the form
\begin{equation}\label{form:fullscal}
G^{(r)}(x,q) = G^{(reg)}(x,q) + \sum_{l=0}^L (1-q)^{\theta_l}
H^{(r)}_l\left(\frac{x_c-x}{(1-q)^\phi}\right) \qquad (x,q) \to (x_c,1).
\end{equation}
is assumed with exponents $\theta_{l+1}>\theta_l$, then by the arguments of the 
last section, the limit $q\to 1$ constrains the exponents to
\begin{equation}
\gamma_{k,l}=\frac{k-\theta_l}{\phi}.
\end{equation}
Comparison with (\ref{form:generexp}) then yields $\theta_l=(l+1)/3$. 
The limit $x\to x_c$ in (\ref{form:fullscal}) provides an expansion of the 
perimeter moments of the form
\begin{equation}
h_k^{(r),(sing)}(q) \sim \sum_{l=0}^L \frac{e_{k,l}}{(1-q)^{\beta_{k,l}}} \qquad (q\to 1),
\end{equation}
where $\beta_{k,l}=k\phi-\theta_l$. For the exponents describing the growth of the 
corresponding series coefficients $a_n=[q^n]h^{(r)}_k(q)$ in (\ref{form:fitass}),
(where $[x^n]g(x)$ denotes the coefficient of $x^n$ in the expansion of the function
$g(x)$), it follows that $\alpha=k\phi$ and $\alpha_i=\theta_i$.
We remark that the number of coefficients $M_0$ used in the fit is much smaller 
than that for the area moments \cite{J03}, where $8\le M_0\le 12$. Apparently, 
the convergence of the perimeter moments to the asymptotic regime is quite slow.
We were initially concerned that this significantly slower convergence was
indicative of some feature of the scaling function we had overlooked. We were
reassured that that is not the case, by performing the same analysis {\em mutatis mutandis}
of the perimeter moment amplitudes for the (exactly solvable) model of staircase polygons.
Precisely the same phenomenon was observed there, and in that case the scaling form
has been proved \cite{P95}.

For given $M$, the amplitude estimates $\{d_i\}$ of (\ref{form:fit}) display cyclic 
fluctuations in $N$. In order to enhance convergence, we considered only
every $r$-th data value, i.e., we determined the coefficients $d_i$ using sets
of equations parametrised by $n=N-r(M+1), N-rM,\ldots, N$ in (\ref{form:fit}),
where $r=2$ for the square and hexagonal lattices, and $r=3$ for the triangular 
lattice. The results of the fit are shown in Table~\ref{tab:unperamp}.
\begin{table}
\begin{center}
\begin{tabular}{ccccc}
Amplitude & Exact value & Square & Hexagonal & Triangular\\ \hline
$e_0$ & unknown  & $-0.3942(2)$ & $-0.6790(4)$  & $-0.476(1)$\\
$e_1$ & unknown &  $-2.576(1)$ & $-5.356(2)$ & $-2.95(1)$ \\
$e_2e_1^{-2}e_0$ & $-0.29052826$ & $-0.29052(2)$ & $-0.29054(5)$ & $-0.2906(1)$\\
$e_3e_1^{-3}e_0^2$ & $0.1396478$ & $0.13967(4)$ & $0.13962(4)$ & $0.13965(3)$\\
$e_4e_1^{-4}e_0^3$ & $-0.0754402$ & $-0.07545(9)$& $-0.07542(8)$ & $-0.0754(1)$\\
$e_5e_1^{-5}e_0^4$ & $0.042564$ & $0.04259(7)$ & $0.0426(1)$ & $0.0426(1)$\\
$e_6e_1^{-6}e_0^5$ & $-0.02448$ & $-0.02451(7)$ & $-0.02446(8)$ & $-0.0245(2)$\\
$e_7e_1^{-7}e_0^6$ & $0.0142143$ & $0.01423(8)$ & $0.01423(9)$ & $0.0144(3)$\\
$e_8e_1^{-8}e_0^7$ & $-0.008292$& $-0.00829(9)$ & $-0.00831(6)$ & $-0.0082(4)$\\
$e_9e_1^{-9}e_0^{8}$ & $0.0048499$& $0.0048(1)$ & $0.00486(8)$ & $0.0048(4)$\\
$e_{10}e_1^{-10}e_0^{9}$ & $-0.0028406$& $-0.0028(2)$ & $-0.00284(7)$ & $-0.0029(3)$\\
\end{tabular}
\end{center}
\caption{\label{tab:unperamp} Bicritical perimeter moment amplitudes of rooted 
self-avoiding polygons}
\end{table}

For the coefficients $e_0$ and $e_1$, the scaling assumption leads to
a prediction in terms of $f_0$ and $f_1$ from (\ref{form:par}).
We get from (\ref{form:par}), on the square lattice, $e_0=-0.3941877(3)$ and 
$e_1=-2.575656(2)$. This agrees, within numerical accuracy, with the estimates 
obtained in Table \ref{tab:unperamp}. Similarly, for the hexagonal lattice, the 
estimates are consistent with the scaling function predictions $e_0=-0.679256(2)$
and $e_1=-5.35661(1)$. For the triangular lattice, we get $e_0=-0.476162(2)$ and 
$e_1=-2.95663(2)$, which is again consistent with the result in Table 
\ref{tab:unperamp}. 

\begin{figure}
\begin{center}
\includegraphics{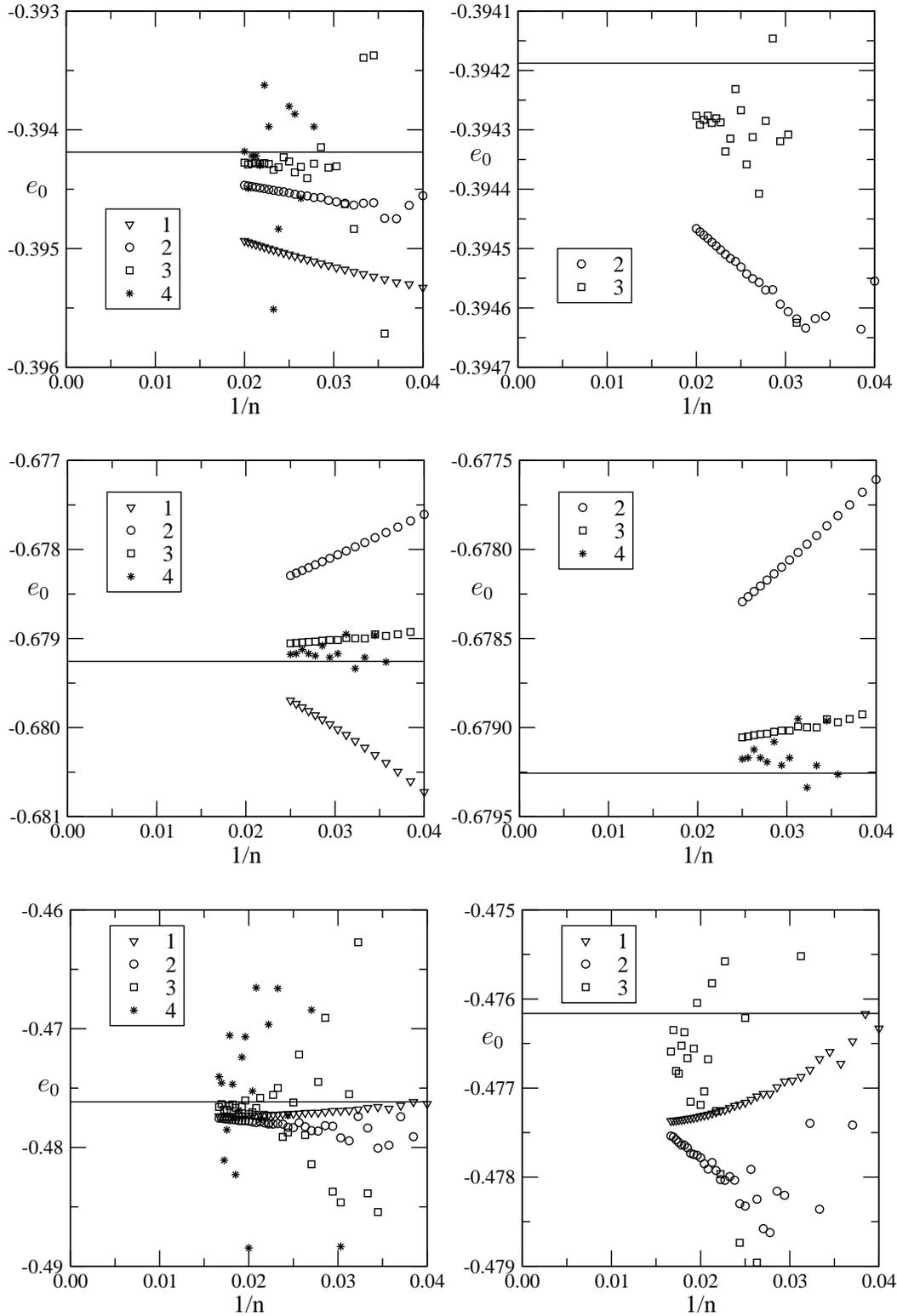}
\end{center}
\caption{\label{fig:ampl} 
Estimates of the amplitude $e_0$ vs. $1/n$ for the square lattice (top panels), 
hexagonal lattice (middle panels) and triangular lattice (bottom panels).}
\end{figure}

It is often useful to check the behaviour of the amplitude 
estimates by plotting the results for the leading amplitude vs. $1/n$. 
In Fig.~\ref{fig:ampl} we have done so for the amplitude $e_0$ for the
square, hexagonal and triangular lattices (the straight lines are the estimates
given above). In the left panels we plot the estimates obtained with
$M$ ranging from 1 to 4 while the right panels give a closer look at
the best converged sequences of amplitude estimates. In each case we
use fits with $\alpha = -1/3$ and $\alpha_i = (i+1)/3$ and as discussed above
we have tried to minimise cyclic fluctuations. We observe that fits with
$M=1$ display pronounced curvature indicating that using just 1 sub-leading 
term gives an insufficient approximation. For the square and triangular
cases the fits with $M=4$ are marred by large fluctuations and are not
very useful. The remaining fits clearly yield estimates for $e_0$ 
fully consistent with the precise values obtained above using the estimates
for $f_0$ and $f_1$. We notice that as more terms are added to the 
fits the estimates exhibits less curvature and that the slope become less 
steep (this is particularly so in the hexagonal case). This is evidence that we 
are indeed fitting to the correct asymptotic form. 

We finally considered the amplitude combinations 
$e_k e_1^{-k} e_0^{k-1}$, where $2\le k\le 10$. They were estimated from the 
ratios
\begin{equation}
\frac{\Gamma(\beta_k)[q^n]h^{(r)}_k(q)\left(\Gamma(\beta_0)[q^n]h_0^{(r)}(q)\right)^{k-1}}
{\left(\Gamma(\beta_1)[q^n]h_1^{(r)}(q)\right)^k} \sim e_k e_1^{-k} e_0^{k-1}
\qquad (n\to\infty).
\end{equation}
We extracted the amplitudes by a direct fit to the expected 
asymptotic form, as explained in the appendix. As argued above, we fitted with 
exponents of the form $\alpha_i=(i+1)/3$ for $1\le i\le M_0$, where 
$2\le M_0\le4$.  The result is shown in Table \ref{tab:unperamp}. The prediction 
of the amplitude combinations appears to be correct, within numerical accuracy.

\section{Unrooted self-avoiding polygons}

The $k$th perimeter moment of rooted self-avoiding polygons $h_k^{(r)}(q)$ is 
related to the $(k+1)$th perimeter moment $h_{k+1}(q)$ of unrooted 
self-avoiding polygons.  We have, for $k\ge0$,
\begin{equation}\label{unroot}
\begin{split}
h_k^{(r)}(q) &= \left.\frac{(-1)^k}{k!} \frac{d^k}{dx^k}
G^{(r)}(x,q)\right|_{x=x_c}=\left.\frac{(-1)^k}{k!} \frac{d^k}{dx^k}
\left( x\frac{d}{dx}G(x,q)\right)\right|_{x=x_c}\\
&= \left.\frac{(-1)^k}{k!}
\left(k \frac{d^k}{dx^k} G(x,q)+x  \frac{d^{k+1}}{dx^{k+1}} G(x,q)
\right)\right|_{x=x_c}\\
&= k\,h_k(q) - (k+1)\,x_c\,h_{k+1}(q) \sim 
- (k+1)\,x_c\,h_{k+1}(q) \qquad (q\to 1)
\end{split}
\end{equation}
The last relation follows with the exponents $\beta_k=2k/3-1/3$ of $h_k^{(r)}(q)$
given in (\ref{form:permomas}). It follows from (\ref{unroot}) that, for $k>0$, the 
singular behaviour of $h_k(q)$ is determined by the singular behaviour of $h_{k-1}^{(r)}(q)$. 
So we have all the moments of unrooted SAP except the zeroth moment. It thus remains to 
analyse the moment $h_0(q)$.
 
Extrapolating the values of $\beta_k$ to $k=-1$ gives $\beta_{-1} = -1$, so we expect a 
singularity of the form
\begin{equation}
h_0^{(sing)}(q) \sim A (1-q)\log (1-q) \qquad(q\to1),
\end{equation}
with some amplitude $A>0$.  This behaviour was tested by a direct fit to the expected 
asymptotic form. Using the notation of the appendix, we expect an exponent $\alpha=-1$ 
and choose $\alpha_i=i$ for $1\le i\le M_0$, where $2\le M_0\le4$ for stable 
approximation schemes.

For the square lattice, we find $c_0=0.026527(6)$. The 
numerical analysis yields $c_0=0.026527(3)$ on the hexagonal lattice, and 
$c_0=0.05306(5)$ on the triangular lattice. This suggests a universal law of the form
\begin{equation}
\sum_m p_{m,n} x_c^{m} \sim \frac{1}{6\pi\sigma} \frac{1}{n^2} \qquad (n\to\infty),
\end{equation}
where $\sigma=2$ for the square lattice and the hexagonal lattice, 
and $\sigma=1$ for the triangular lattice.
We have $1/(12\pi)=0.0265258...$ and $1/(6\pi)=0.0530516...$, which
is, within error bars, in agreement with the estimates obtained above.

If one accepts the predicted scaling form (\ref{form:scal}) and scaling function
(\ref{form:sfun}) of rooted
self-avoiding polygons, (and we believe that we have provided compelling numerical
evidence to do so), then the scaling function of (unrooted) self-avoiding polygons
is determined by integration,
\begin{equation}\label{form:newscal}
G^{(sing)}(x,q) \sim (1-q) F\left(\frac{x_c-x}{(1-q)^{2/3}}\right) + C(q), 
\qquad (x,q)\to (x_c,1),
\end{equation}
where $C(q)$ is a ``constant'' of integration, given below, and $F(s)$ is given by
\begin{equation}\label{ursf}
F(s)=\frac{4f_1}{ x_c} \log \mbox{Ai}\left( \left(\frac{f_0}{4f_1}\right)^{2/3}s\right),
\end{equation}
where the coefficients $f_0$ and $f_1$ are given in Table \ref{tab:amp}, and 
$x_c$ is given in the introduction.
Alternatively, $F(s)$ can be expressed in terms of the amplitude $E_0$ given in
Table \ref{tab:amp} by
\begin{equation}\label{form:oldsf}
F(s)=-\frac{1}{\pi\sigma} \log \mbox{Ai}\left( \frac{\pi}{x_c}(2E_0)^{2/3}s\right).
\end{equation}
The term $C(q)$ incorporates the singular behaviour of the bicritical area generating function
$h_0(q)=G(x_c,q)$ as $q$ approaches unity, since the limit $x\to x_c$ from the other 
term yields a vanishing contribution. The function $C(q)$ is thus given by
\begin{equation}\label{form:add}
C(q)=\frac{1}{6\pi\sigma}(1-q)\log(1-q).
\end{equation}
The scaling form (\ref{form:newscal}), (\ref{form:oldsf}) and (\ref{form:add}) 
refines the prediction given previously in \cite{RGJ01} by the addition of the term 
(\ref{form:add}).

\section{Crossover to branched polymer phase}

Let $x_c(q)$ denote the radius of convergence of the perimeter and area generating function 
$G^{(r)}(x,q)$ of rooted SAP for $q<1$ fixed, as $q\to1$. The scaling function prediction 
(\ref{form:scal}) leads to a prediction of the slope of the critical line $x_c(q)$, see also 
\cite{C01,R02}. The slope of the critical line is determined by the first singularity 
$s_c$ of the scaling function $H^{(r)}(s)$ on the negative real axis. More precisely, for 
$q<1$ fixed, the argument $s=(x_c-x)/(1-q)^\phi$ of the scaling function is negative for 
$x>x_c$, attaining its singular value $s_c$ for $x=x_c(q)>x_c$. We thus expect asymptotically
\begin{equation}
x_c(q)\sim x_c-s_c(1-q)^\phi \qquad (q\to1).
\end{equation}
For the particular scaling function (\ref{form:sfun}), the point $s_c$ is given by
\begin{equation}
\left(\frac{f_0}{4f_1}\right)^{\frac{2}{3}} s_c = -2.3381074104\ldots
\end{equation}
From the values of Table \ref{tab:amp}, we obtain $s_c=-0.2608637(5)$, $-0.2161405(20)$, 
and $s_c=-0.274509(2)$ for the square, hexagonal, and triangular lattices respectively. 
Note that, due to the particular form of the scaling function, the same behaviour applies 
to unrooted self-avoiding polygons.

\begin{figure}
\begin{center}
\includegraphics{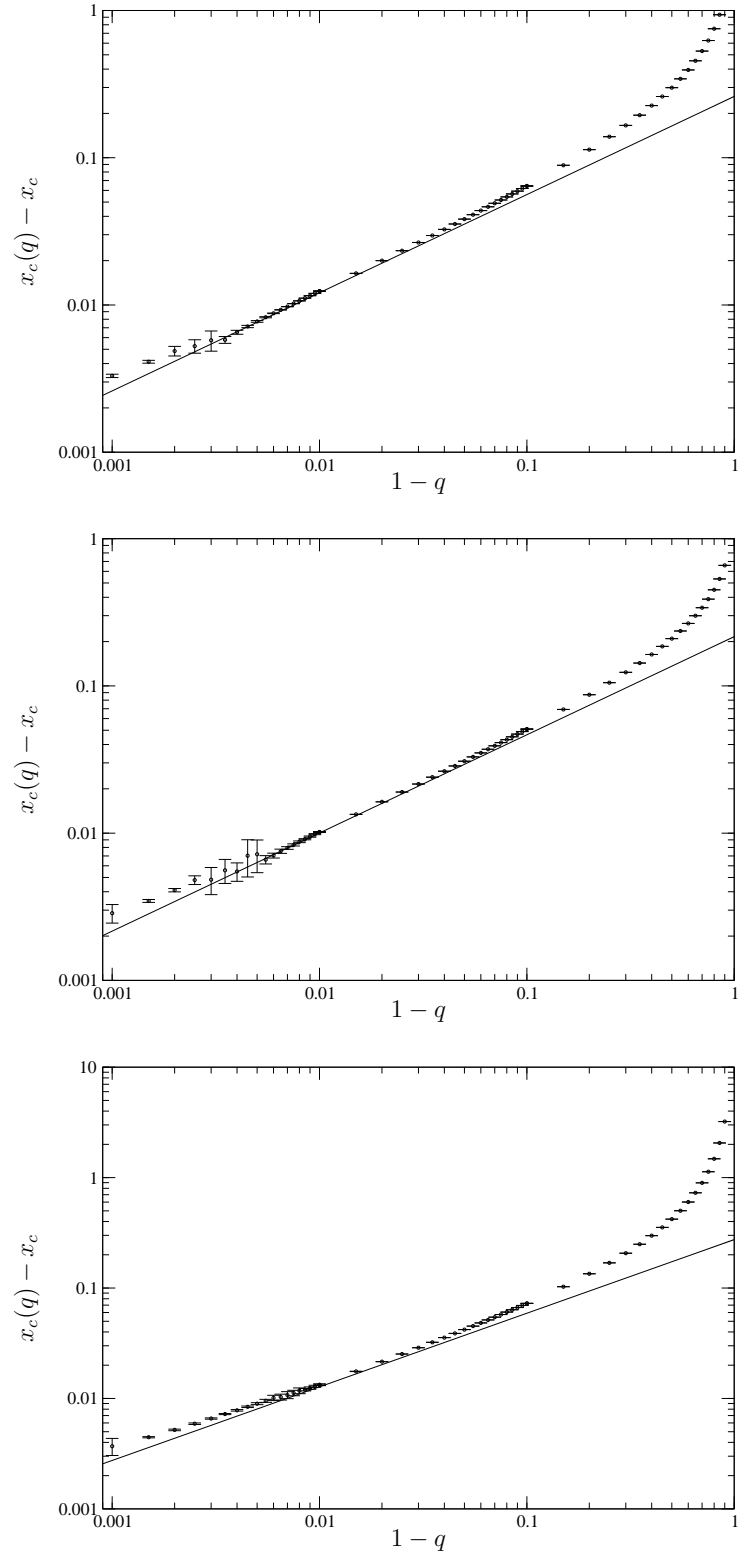}
\end{center}
\caption{\label{fig:cross} 
Plots of $x_c(q)-x_c$ versus $1-q$ for the square lattice (top), 
hexagonal lattice (middle) and triangular lattice (bottom). The solid line
has the predicted slope $2/3$.}
\end{figure}

Figure \ref{fig:cross} displays a log-log plot of $x_c(q)-x_c$ versus $1-q$. 
In each plot a straight line corresponding to the expected form
\begin{equation}
x_c(q)-x_c = -s_c (1-q)^{2/3}
\end{equation}
is given. We get reasonable agreement with the predicted form. 
The estimates were 
obtained using third order differential approximants \cite{G89}, with the degree of 
inhomogeneous polynomial ranging from 5 to 15, and the requirement that averages 
must include at least $85\%$ of the approximants. 
For this part of our study we calculated the numbers $p_{m,n}$ for $m\le 90$ (square lattice), 
$m\le 134$ (hexagonal lattice) and $m\le 50$ (triangular lattice), and all relevant
sizes of the area. 

If the scaling function $H^{(r)}(s)\sim a(s-s_c)^{g}$ about $s=s_c$, it follows that the singular 
part of $G^{(r)}(x,q)$ behaves as $G^{(r),(sing)}(x,q) \sim a(1-q)^{-g\phi}(x_c(q)-x_c)^g$
for $x\to x_c(q)$. The scaling function (\ref{form:sfun}) has a simple pole at $s_c$ so
\begin{equation}
G^{(r),(sing)}(x,q) \sim a(1-q)^{2/3}(x_c(q)-x_c)^{-1}.
\end{equation}
The differential approximant analysis typically yields reasonably accurate exponent estimates for 
$q\leq 0.995$ and confirms that $G^{(r)}(x,q)$ has a simple pole as $x\to x_c(q)$.
For example the analysis of the square lattice series yields $g=-1.0(3)$ at $q=0.995$,
$g=-1.03(2)$ at $q=0.99$ and $g=-0.999(2)$ at $q=0.95$. For $q$ closer to 1 the exponent
estimates became unreliable in that the errors bars were as large as the estimates, 
e.g. at $q=0.997$ we found $g=-1.6 \pm 1.7$.
We also tried to calculate the amplitude $a(q)\sim a(1-q)^{2/3}$, but unfortunately we could
not get accurate estimates for $q>0.9$, and so have been unable to  numerically confirm
the predicted behaviour. 

\begin{figure}
\begin{center}
\includegraphics{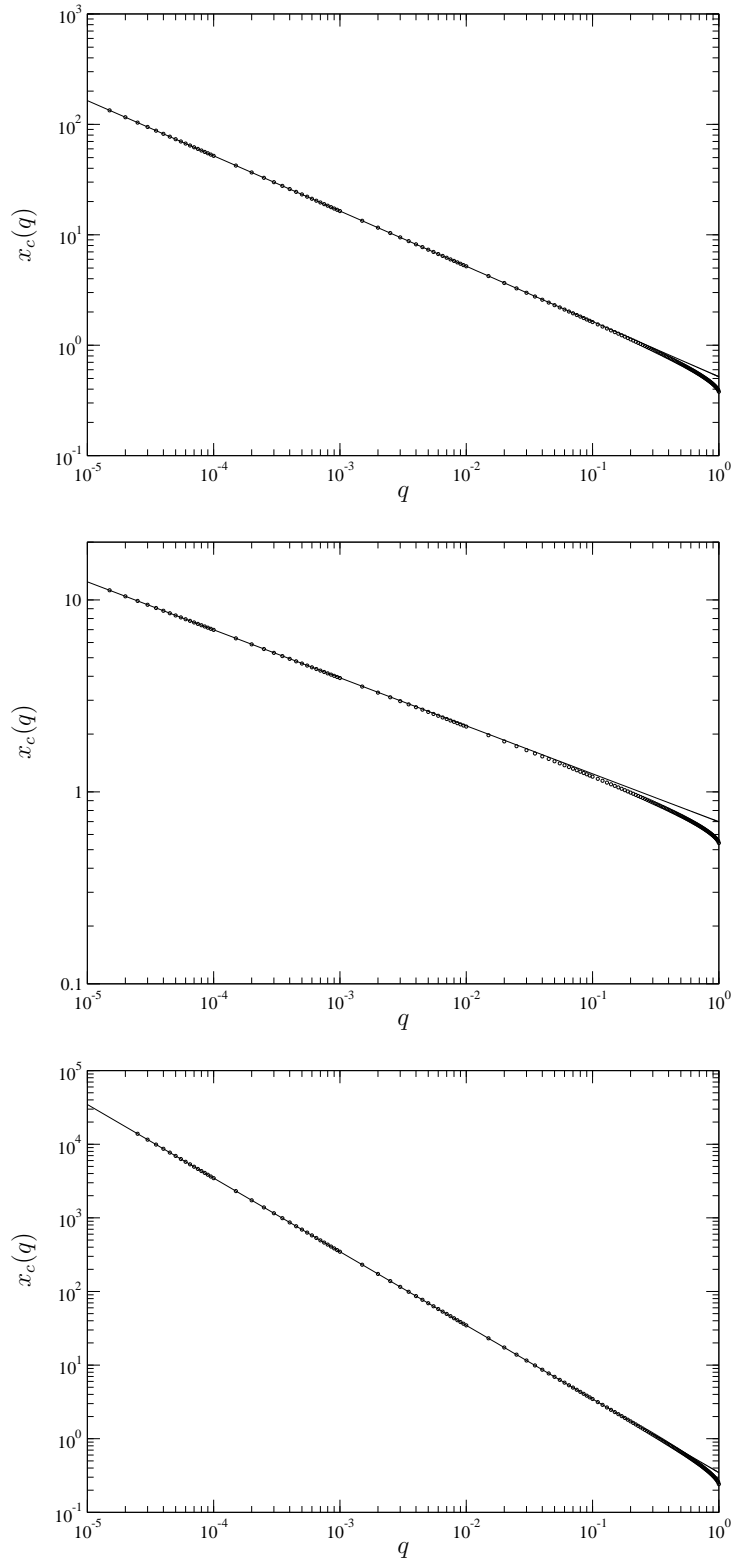}
\end{center}
\caption{\label{fig:qto0} 
Plots of $x_c(q)$ versus $q$ for the square lattice (top), hexagonal lattice 
(middle) and triangular lattice (bottom). The solid line has the predicted 
form $x_c(q) = t_c/q^{b}$  with slope $b=1/2$, $1/4$ and 1, respectively.}
\end{figure}

Finally, we checked the behaviour of the phase boundary $x_c(q)$ as $q\to 0$. 
The coefficient of $x^m$ in $G^{(r)}(x,q)$ is a polynomial in $q$ and as $q\to 0$
it becomes completely dominated by the term of lowest degree in $q$.  
We thus have to examine the behaviour of the terms $a_m =p_{m,n_{\rm min}}$, where
$p_{m,n_{\rm min}}$ is the number of polygons with perimeter $m$ having the {\em minimal} 
possible area $n_{\rm min}$. Clearly, the polygon formed by making a linear chain
of unit cells contributes to $a_m$. Unit cells on the square, hexagonal and
triangular lattices have perimeter 4, 6 and 3, respectively, and it thus follows that 
$n_{\rm min} \simeq m/2$ (square), $m/4$ (hexagonal) and $m$ (triangular).
On the square lattice it has been shown \cite{FGW} that $a_m$ grows exponentially with $m$.
$a_m$ is bounded from above by the number of site trees of size $n_{\rm min}$
on the dual lattice and from below by the number of minimally spanning polyominoes  
of size $n_{\rm min}$ (a minimally spanning polyomino is a polyomino spanning
a $h\times w$ rectangle having size $h+w+1$). Similar arguments apply to the other
lattices. So we form the generating function $S(t) = \sum_m a_m t^m$ and using 
differential approximants find that $S(t)$ has a singularity at $t_c=0.5189688(2)$ on the 
square lattice, $t_c=0.6986253(5)$ on the hexagonal lattice, and $t_c=0.346530(1)$ on the 
triangular lattice.  Since $t^m = x^m q^{n_{\rm min}}$ it follows that $x_c(q) \sim t_c/q^b$, 
with $b=1/2$, $1/4$ and  $1$ for the square, hexagonal and triangular lattice, respectively.
Figure \ref{fig:qto0} shows a log-log plot of $x_c(q)$ versus $q$. 
In each plot a straight line corresponding to the expected form
$x_c(q) = t_c/q^{b}$ is also shown.

\section{Conclusions}

We have analysed bicritical perimeter moments of self-avoiding polygons using data obtained from
exact enumeration on the square, hexagonal and triangular lattices. This yields a new check of 
the earlier scaling function conjecture for self-avoiding polygons. The numerical analysis 
supports the crossover behaviour of the critical line to the branched polymer regime. Whereas 
we find the scaling function conjecture for rooted self-avoiding polygons satisfied, it can be
valid for unrooted self-avoiding polygons only in modified form. By analysing the bicritical
area generating function, we suggest a modification by an additional term 
with an apparently universal amplitude, see (\ref{form:newscal}), (\ref{form:oldsf}) 
and (\ref{form:add}). It would be interesting to consider whether its value can be 
justified by field theoretical arguments; compare the related investigation 
\cite{CZ02}. 

\section*{Acknowledgements}

CR would like to acknowledge funding by the German Research Council (DFG). He thanks
the Department of Mathematics and Statistics for hospitality and partial funding of
a stay at Melbourne University in winter 2003, where part of this work was done.
IJ and AJG are happy to acknowledge financial support from 
the Australian Research Council. 
IJ gratefully acknowledges a generous grant of computing resources from the
Australian Partnership for Advanced Computing (APAC) without which 
the numerical calculations presented in this paper would have been impossible. 
We also gratefully acknowledge use of the computational resources of the Victorian 
Partnership for Advanced Computing (VPAC).

\section*{Appendix: Numerical methods}

We numerically analyse sequences $a_n$ by a direct fit to
the expected asymptotic form. Similar applications of this method can be 
found in \cite{JG99,J03}. The sequence $a_n$ is assumed to behave 
asymptotically as
\begin{equation}\label{form:fitass}
a_n \sim \mu^{n} n^{\alpha-1} \left( c_0
+\frac{c_1}{n^{\alpha_1}}+\frac{c_2}{n^{\alpha_2}}+\ldots+\frac{c_M}{n^{\alpha_M}}\right)\qquad
(n\to\infty),
\end{equation}
with constants $\mu>0$, $c_i\neq0$ for $i=0,1,\ldots,M$, and exponents $\alpha$ and 
$\alpha_i$ for $i=1,\ldots,M$, where $\alpha_{i+1}>\alpha_i$ for $i=1,\ldots, M-1$.
Estimates of the constant $\mu$ and the exponent $\alpha$ can be obtained by, e.g., the
method of differential approximants \cite{G89}. We have $\mu=1$ in our examples.
Often the sequence $(\alpha_i)_{i=1}^M$ is unknown, but there are predictions for 
the numbers $\alpha_i$. The validity of a prediction can be tested employing the 
following procedure:

We perform a direct fit to the expected asymptotic form, i.e., we solve the linear system
\begin{equation}\label{form:fit}
a_n = \mu^n n^{\alpha-1} \left( d_0
+\frac{d_1}{n^{\alpha_1}}+\frac{d_2}{n^{\alpha_2}}+\cdots+\frac{d_M}{n^{\alpha_M}}\right)
\qquad (n=N-(M+1), \ldots,N)
\end{equation}
for the $M+1$ unknowns $d_i$.  If the assumption of the asymptotic form (\ref{form:fitass})
is correct, then the numbers $d_i=d_i(N,M)$ will satisfy
\begin{equation}
d_i(N,M) \to c_i \qquad (N\to\infty).
\end{equation}
Generally, if the wrong sequence has been chosen, the sequence of 
coefficients $d_i(N,M)$, for increasing values of $N$ and fixed $M$, diverges either to 
infinity or converges to zero; but if the correct sequence has been chosen, convergence is 
usually rapid and obvious.

Let us fix $i$ in the following. Estimates of the amplitudes $c_i$ are obtained in the following way. 
For $N$ large enough, the numbers $d_i(N,M)$ display approximately linear variation in $1/N$,
\begin{equation}\label{form:asreg}
d_i(N,M) \sim c_i(M) + \frac{r_i(M)}{N} \qquad (N\to\infty).
\end{equation}
The numbers $c_i(M)$ are obtained by a linear least squares fit of $d_i(N,M)$ for large values 
of $N$. The larger $M$, the larger $N$ has to be taken in order to reach the 
asymptotic regime (\ref{form:asreg}). Since, however, $N\le N_0$ for given series data,
$d_i(N,M)$ is close to the asymptotic regime only for values of $M\le M_0$, where $M_0$ has to
be extracted from series analysis. We choose $1\le m_0\le M_0$ such that $|r_i(m_0)|$ is minimal.
We estimate $c_i$ by $c_i(m_0)$ and estimate the error by the spread among different values
$c_i(m)$ for $1\le m\le M_0$.

The amplitudes $c_i$ in (\ref{form:fitass}) are related to the critical amplitudes of the 
corresponding generating functions. If $\alpha\notin -\mathbb N_0$, the singular amplitude $B$ 
of the corresponding generating function 
\begin{equation}
A(x)=\sum_{n=0}^\infty a_n x^n \sim A^{(reg)}(x)+\frac{B}{(\mu^{-1}-x)^\alpha} \qquad (\mu x\to1)
\end{equation}
is related to $c_0$ via $B = c_0\Gamma(\alpha)\mu^{-\alpha}$.

\end{document}